\documentstyle{mn}
\def\gapprox{\mathrel{\vcenter{\offinterlineskip \hbox{$>$}
    \kern 0.3ex \hbox{$\sim$}}}}
\def\lapprox{\mathrel{\vcenter{\offinterlineskip \hbox{$<$}
    \kern 0.3ex \hbox{$\sim$}}}}

\title{Magnetohydrodynamical Non-radiative Accretion Flows in Two-Dimensions}

\author[]{James M. Stone$^{1}$\thanks{E-mail: jstone@astro.umd.edu}
and James E. Pringle$^{2}$\thanks{E-mail: jep@ast.cam.ac.uk} \\
$\rm ^1$ Department of Astronomy, University of Maryland, College Park, MD
20742-2421 USA \\
$\rm ^2$ Institute of Astronomy, Cambridge University, Madingley Road,
Cambridge CB3 0HA, UK \\}

\begin{document}
\maketitle
\begin{abstract}
We present the results of axisymmetric, time-dependent magnetohydrodynamic
simulations of accretion flows around black holes.  The calculations 
begin from a rotationally supported thick torus which contains a
weak poloidal field.  Accretion is produced by growth and saturation
of the magnetorotational instability (MRI) provided the wavelength of
the fastest growing mode is less than the thickness of the torus.
Using a computational grid that spans more than two decades in radius,
we compare the time-averaged properties of the flow to previous
hydrodynamical simulations.  The net mass accretion rate is small
compared to the mass inflow and outflow rates at large radii associated
with turbulent eddies.  Turbulence is driven by the MRI rather than
convection.  The two-dimensional structure of the time-averaged flow is
significantly different compared to the hydrodynamical case.  We
discuss the limitations imposed on our results by the assumption of
axisymmetry and the relatively small radial domain.

\end{abstract}


\section{Introduction}

Low-luminosity accretion flows onto compact objects are generally thought to be
non-radiative, i.e. they cannot lose internal energy through radiative
cooling.  The structure of such flows is a subject of great interest,
and has been studied both through self-similar solutions assuming
steady-state (e.g. Narayan \& Yi 1994; 1995 and references therein;
Abramowicz et al.  1995; Blandford \& Begelman 1999) and more recently,
through direct numerical hydrodynamical simulations (Igumenshchev \&
Abramowicz 1999, hereafter IA99; Stone, Pringle, \& Begelman 1999,
hereafter SPB; Igumenshchev \& Abramowicz 2000).

Numerical solutions to the hydrodynamic equations offer a new
opportunity to study the multidimensional and time-dependent nature of
non-radiative accretion flows.  In a purely hydrodynamical flow
accretion must be driven by the addition of an anomalous shear stress
to the equations of motion.  The form and amplitude of this stress is
arbitrary; often it is assumed to take the form of a kinematic
viscosity with coefficient $\nu = \alpha C_{s}^{2}/\Omega$, where $\alpha$ is a
dimensionless constant, $C_{s}$ the sound speed, and $\Omega$ the orbital
frequency (Shakura \&
Sunyaev 1973).  Despite this limitation, the simulations reveal a
variety of interesting characteristics of such flows.  For example, SPB
found that, regardless of the amplitude of the anomalous shear stress,
the flow at small radii is dominated by strong radial convective
motions.  The mass flux associated with the convective eddies is very
large compared to the net mass accretion rate through the disks, which
is determined by the properties of the accretion flow at the inner
boundary.  For small values of the anomalous shear stress, these
results were in good agreement with earlier work by IA99.  However, for
very large shear stresses, IA99 found that convection could be
suppressed, and a more laminar inflow/outflow solution was possible.
These differences appear to be sensitive to the details of the assumed
anomalous stress (Miller \& Stone 2000).  In particular, if all
components of the viscous stress tensor are included in the equations
of motion (so that the stress represents a true kinematic viscosity),
then convection is suppressed at large values of $\alpha$ since the
Reynolds number of the flow is then very small.  On the other hand, if
only the azimuthal components of the stress are included (in order to
represent the highly anisotropic nature of the shear stress generated
by magnetohydrodynamic turbulence, see Balbus \& Hawley 1998), then
strongly buoyant convective bubbles are observed even at large values
of the shear stress.

The vanishingly small accretion rates observed at small $\alpha$ in
both IA99 and SPB have been interpreted as due to inward angular
momentum transport in axisymmetric convective flows.  Recently,
self-similar solutions which reproduce the properties of the
simulations have been described by Narayan, Abramowicz, \& Igumenshchev
(2000) and Quataert \& Gruzinov (2000).  These authors argue that in
three-dimensions (3D), turbulent mixing of convective eddies might transport
angular momentum outwards, and so change the qualitative nature of the
flow (although 3D direct numerical simulations of turbulent convection in
{\em thin} accretion disks have shown that in this case, angular
momentum is not carried outwards: Stone \& Balbus 1996; Cabot 1996).
Fortunately, fully 3D simulations of
hydrodynamical non-radiative accretion flows that test these ideas are
possible (Igumenshchev, Abramowicz, \& Narayan 2000; Miller \& Stone
2000); it is found that in 3D the flows are
qualitatively and quantitatively similar to the axisymmetric models,
and moreover, that angular momentum transport in thick disks via
radial convection is
vanishingly small, or perhaps even inward.  A kinetic theory for
hydrodynamic turbulence in accretion flows that can account for these
results has been developed by Quataert \& Chiang (2000).
A low mass accretion rate and central density is one possible
interpretation of the low flux of high-energy emission observed
from black holes at the center of early-type galaxies (Di Matteo,
Carilli, \& Fabian 2000), and the possible detection of polarized
emission from the galactic center (Agol 2000; Quataert \& Gruzinov 2000).

While understanding the properties of hydrodynamical accretion flows
driven by an anomalous shear stress is important, it is generally
agreed that angular momentum transport is in fact mediated by magnetic
stresses, either through magnetohydrodynamic (MHD) turbulence driven by
the magnetorotational instability (MRI) (Balbus \& Hawley 1998), or
global stresses associated with a magnetocentrifugal wind (Blandford \&
Payne 1981).  Thus, it is essential to extend the hydrodynamical models
to MHD.  In this paper we present the results of a series of MHD
simulations of non-radiative accretion flows in which accretion is
driven self-consistently by magnetic stresses within the flow.

The evolution of magnetized accretion flows has been studied via
numerical methods in a variety of contexts.  Studies of thin disks
threaded by a vertical field have been reported by a number of authors
(Shibata \& Uchida 1986; Stone \& Norman 1994; Matsumoto et al. 1996;
Kudoh et al. 1998).  Rapid collapse of the disk is observed in these
calculations, driven either by axisymmetric modes of the MRI when the
field is weak, or by magnetic braking effects when the field is
strong.  On the other hand, the nonlinear stage of the MRI has been
studied extensively in local 3D simulations (Hawley, Gammie, \& Balbus
1995; 1996 hereafter HGB; Brandenburg et al.  1995; Stone et al. 1996),
and more recently global 3D simulations (Matsumoto 1999; Armitage 1998;
Hawley 2000).  In this case, MHD turbulence is the inevitable outcome
of the MRI.

The study of Hawley (2000) in particular is relevant to the results
presented here.  He presents a number of time-dependent MHD
simulations of accretion onto a black hole in two- and
three-dimensions, beginning from a pressure
supported torus with a weak, embedded magnetic field.  In each case,
development of the MRI produces MHD turbulence and accretion after only
a few orbits.  Comparison of the evolution of axisymmetric and fully 3D
simulations reveals important differences.  In 3D development of the
nonaxisymmetric MRI increases the level of MHD turbulence and sustains
mass accretion throughout the entire evolution, whereas in the
axisymmetric models accretion dies away on long timescales as there is
no dynamo action to maintain poloidal field.

The results presented in this paper are in many ways complementary to
those of Hawley (2000).  Although we limit this study entirely to
axisymmetric models, this allows us to afford a much larger radial
domain than current 3D models (more than a factor of ten in some
cases).  This introduces a very large range of dynamical timescales
between the inner and outer regions of the disk.  Thus, although
turbulence and accretion inevitably die away in axisymmetry after a few
orbits of the outer disk, the inner disk undergoes thousands of orbits
in this same period.  Thus, the flow may reach a quasi steady-state in
the inner regions, which can be compared and contrasted to the
steady-state structure found for purely hydrodynamical models.  In
fact, we find both important differences and similarities between
axisymmetric hydrodynamical models (e.g. SPB) and the MHD models
described here.  For example, convective instabilities appear to play
little role in determining the two-dimensional structure of the flow,
instead turbulence driven by the MRI dominates.  We find that unbound
outflows can be produced by magnetic stresses in models with strong
poloidal fields.  In every case we have studied here, the mass
accretion rate is once again small (compared to the expectations of
some self-similar steady-state solutions such as Advection Dominated
Accretions Flows or ADAFs; Narayan \& Yi 1995), and determined by
the flow properties at the inner boundary.  It will, however, be
important to test these results with three-dimensional simulations that
span a large radial domain.

The paper is organized as follows.  In \S2 we describe our methods, in
\S3 we present our results, in \S4 we discuss these results, and in \S5
we summarize and conclude.

\section{Method}

\subsection{The Equations of Motion}

To compute the models discussed here, we solve the equations of MHD
\begin{equation}
 \frac{d\rho}{dt} + \rho \nabla \cdot {\bf v} = 0,
\end{equation}
\begin{equation}
 \rho \frac{d{\bf v}}{dt} = -\nabla P - \rho \nabla \Phi +
 \frac{1}{4\pi}(\nabla \times {\bf B}) \times {\bf B}
\end{equation}
\begin{equation}
 \rho \frac{d(e/\rho)}{dt} = -P\nabla \cdot {\bf v} + \eta {\bf J}^{2}
\end{equation}
\begin{equation}
 \frac{\partial {\bf B}}{\partial t} = \nabla \times ({\bf v} \times {\bf B}
- \eta {\bf J})
\end{equation}
where $\rho$ is the mass density, $P$ the pressure, ${\bf v}$ the
velocity, $e$ the internal energy density, ${\bf B}$ the magnetic
field, and ${\bf J} = (c/4\pi)\nabla \times {\bf B}$ the current density.
Note that the equations of motion do not contain
any anomalous shear or viscous stress terms.  We
adopt an adiabatic equation of state $P=(\gamma -1)e$, and consider
models with $\gamma =5/3$.  These equations are solved in spherical
polar coordinates $(r,\theta,\phi)$.

The final terms in equations [3] and [4] are the magnetic heating and
diffusion rates mediated by a finite resistivity $\eta$.
For numerical schemes that solve the internal (rather
than total) energy equation [3], numerical diffusion of the magnetic
field can result in the loss of energy from the system.  We expect this
effect could be important for the adiabatic and turbulent flows studied
here.  In order to capture this energy in the form of heating at
current sheets, we have added an anomalous resistivity of the form
\begin{equation}
 \eta = \eta_{\circ} | {\bf J} |
\end{equation}
The form for $\eta_{\circ}$ and the implementation in our numerical
code is described in an Appendix.  Comparison of the evolution of
magnetized tori with and without an anomalous resistivity show that in
the former case total energy conservation is improved to better than
one part in a thousand.  Note that by making the anomalous resistivity
proportional to the current density, we ensure it is large in current
sheets (and therefore spreads them out over enough zones to be
resolved), but has negligible effect in smooth regions of the flow.
Thus, the anomalous resistivity adopted here plays a role analogous to
the artificial viscosity used to capture shocks (Stone \& Norman 1992a).

All of the simulations presented here use the pseudo-Newtonian gravitational
potential introduced by Paczy\'{n}ski \& Wiita (1980)
\begin{equation}
 \Phi = - \frac{GM}{r-R_{G}}
\end{equation}
to approximate general relativistic effects in the inner regions.  In
particular, this potential reproduces the last stable circular orbit at
$r=3R_{G}$.  Thus, the flows we study here are
primarily intended to model accretion onto a non-rotating black hole.
We find the use of this potential simplifies the inner boundary
condition for the magnetic field, as discussed below.

\subsection{Initial Conditions}

As in SPB, our simulations begin with an exact equilibrium state
consisting of a constant angular momentum torus embedded in a
non-rotating, low-density ambient medium in hydrostatic equilibrium.
The pressure, density, and angular velocity in the torus initially are
given by an analytic solution generated by the procedure described by
Papaloizou \& Pringle (1984), see SPB for details.  The shape and size
of the torus are controlled by two parameters, the distortion parameter
$d$ and the maximum density $\rho_{max}$, as well as the polytropic
index $n=(\gamma-1)^{-1}$.  We adopt $d=1.5$ and $\rho_{max}=1$ for all
the models computed here.  Note we use a larger value of $d$ than the
models presented in SPB; this gives a thicker torus initially in which
there is a larger dynamic range of unstable modes for the MRI.  The
torus is embedded in an ambient medium with constant density
$\rho_{0}=10^{-4}$ and pressure $P_{0} = \rho_{0}/r$.

The magnetic field which threads the torus initially is generated by a
vector potential, i.e. ${\bf B} = \nabla \times {\bf A}$.  Initializing
the field in this way guarantees it will be divergence free.
We take ${\bf A}$ to be purely azimuthal with
\begin{equation}
 A_{\phi} = \rho^{2} / \beta_{\circ}
\end{equation}
where $\beta_{\circ}$ is an input parameter that specifies the magnetic
field strength.  This form for ${\bf A}$ generates poloidal field loops
that are parallel to the density contours, and with an amplitude such
that the parameter $\beta_{\circ}$ corresponds approximately to the
plasma $\beta$-parameter (where $\beta \equiv 8\pi P / B^{2}$) in the
interior of the torus initially.  Note that in contrast to some earlier
studies of the evolution of magnetized accretion disks (e.g. Shibata \&
Uchida 1986; Stone \& Norman 1994), the magnetic field is initially
confined to the interior of the torus and does not connect to the
ambient medium.

\subsection{Numerical Methods}

All of the calculations presented here use the ZEUS-2D code described
by Stone \& Norman (1992a; 1992b). Several minor modifications to the code 
were required for the calculations presented here, including
implementing the Paczy\'{n}ski \& Wiita potential, and adding
the anomalous resistivity as described in the Appendix.

The pseudo-Newtonian potential introduces a radial scale into the
problem: the radius of the event horizon $R_{G}$.
Our computational grid extends
from an inner boundary at $r=2R_{G}$ to either $160R_{G}$ (with the
density maximum of the torus located initially at $R_{0}=40R_{G}$), or
$400R_{G}$ (with $R_{0}=100R_{G}$).  In the latter case, this gives a
spatial dynamic range over which an accretion flow can be established
of nearly two orders of magnitude.  This range is smaller than that
used by SPB because of the much greater computational cost of
MHD compared to hydrodynamical simulations, primarily caused by the
much smaller timestep enforced by stability requirements in regions of
high Alfv\'{e}n speed.  Some models have been evolved for up to
5 million cycles, taking more than 400 hours on a modern workstation.

As in SPB, we adopt a non-uniform, logarithmically spaced grid, so that
$(\bigtriangleup r)_{i+1} / (\bigtriangleup r)_{i} =
\sqrt[N_{r}-1]{10}$.  This gives a grid in which $\bigtriangleup r
\propto r$, with $N_{r}$ grid points per decade in radius.  Similarly,
we adopt non-uniform angular zones with $(\bigtriangleup \theta)_{j} /
(\bigtriangleup \theta)_{j+1} = \sqrt[N_{\theta}-1]{4}$ for $0 \leq
\theta \leq \pi/2$ (that is the zone spacing is decreasing in this
region), and $(\bigtriangleup \theta)_{j+1} / (\bigtriangleup
\theta)_{j} = \sqrt[N_{\theta}-1]{4}$ for $\pi/2 \leq \theta \leq
\pi$.  This gives a refinement by a factor of four in the angular grid
zones between the poles and equator.  Equatorial symmetry is not
assumed.  Our standard resolution is $N_{r}=64$ and $N_{\theta}=44$.
We have also computed a high resolution model with $N_{r}=128$ and
$N_{\theta}=80$.

We adopt outflow boundary conditions (projection of all dynamical
variables except the magnetic field) at both the inner and outer radial
boundaries.  For the magnetic field at the outer radial boundary we use
an outflow condition (projection of each component), whereas at the
inner boundary we use a negative-stress condition (that is we enforce
$B_{r}B_{\phi} \leq 0$ at $r=2R_{G}$).  This ensures there is no outward
flux of energy or angular momentum at the inner boundary.  Our radial
grid is large enough that the outer boundary condition has little
effect.  Moreover, the use of the pseudo-Newtonian potential greatly
simplifies the inner boundary condition provided the inner edge of the
grid is well within the last stable orbit, because in this case
a super-Alfv\'{e}nic
inflow is always present there.  In the angular direction,
the boundary conditions are set by symmetry at the poles.

\begin{table*}
\footnotesize
\caption{Properties of Simulations}
\begin{tabular}{cccccccc} \\ \hline
Run
& $\beta_{\circ}$
& Resolution$^{(a)}$
& $R_{0}/R_{G}$
& $t_{f}$ (orbits$^{(b)}$) \\ \hline

A & 100      &  64 & 40 & 5.08  \\
B & 200      &  64 & 40 & 5.82  \\
C & 400      & 128 & 40 & 5.41  \\
D & $\infty$ &  64 & 40 & 5.00  \\
E & 100      &  32 & 100 & 5.25  \\
F & 200      &  64 & 100 & 4.50  \\
G & $\infty$ &  64 & 100 & 4.00  \\
\hline
\end{tabular}\\
(a) grid points per decade in radius, (b) at $r=R_{0}$\\
\end{table*}

\section{Results}

\subsection{A Fiducial Model}

Table 1 summarizes the properties of the simulations discussed here.
Columns two through five give the initial field strength (as measured
by $\beta_{\circ}$), the numerical resolution, the radial extent of the
numerical grid expressed as the ratio $R_{0}/R_{G}$, and the
final time $t_{f}$ at which each simulation is stopped (all times in
this paper are reported in units of the orbital time at $r=R_{0}$).

Run C has the highest resolution and weakest initial magnetic field,
and therefore the largest range in unstable wavelengths of the MRI: we
discuss it first (we find the fastest growing mode of the MRI
is unresolved in this case at lower resolution, see below).  Figure~1
plots the time evolution of the density in Run C.  In each panel, gray
shading is used to denote regions of the torus in which $v_{r} > 0$.
The growth rate of the MRI is highest near the inner edge of the torus
(since the angular frequency $\Omega$ is largest there), thus
saturation of the MRI produces inward mass flow from the inner edge of
the torus first.  In local axisymmetric simulations of the MRI with
initially uniform vertical fields (Hawley \& Balbus 1992), an organized
flow consisting of two oppositely directed radial streams called the
channel solution emerges in the nonlinear regime.  There is no evidence
that the channel solution persists over a large range of radii in these
global models; such solutions are found to dominate axisymmetric
simulations of thin disks (e.g. Stone \& Norman 1994).  This may be a
result of the radial variation in density and magnetic field within the
torus, or because of the rapid change in timescales with radius, so
that channels that form first at the inner edge quickly encounter
slower material as they move outwards.  Accretion through the inner
boundary begins between 1 and 2 orbits.  Thereafter, the accretion flow
thickens considerably.  Wiggles in the contours indicate the presence
of large amplitude density fluctuations, and moreover the complex
pattern of shading indicates both inward and outward moving fluid
elements are located at all angles.  Overall, the flow has many
similarities to the hydrodynamical torus (see Figure 1 of SPB),
although in this case the fluctuations in the flow are not driven by
convection but rather the MRI.

\begin{figure*}
\begin{picture}(504,360)
\put(0,0){\includegraphics{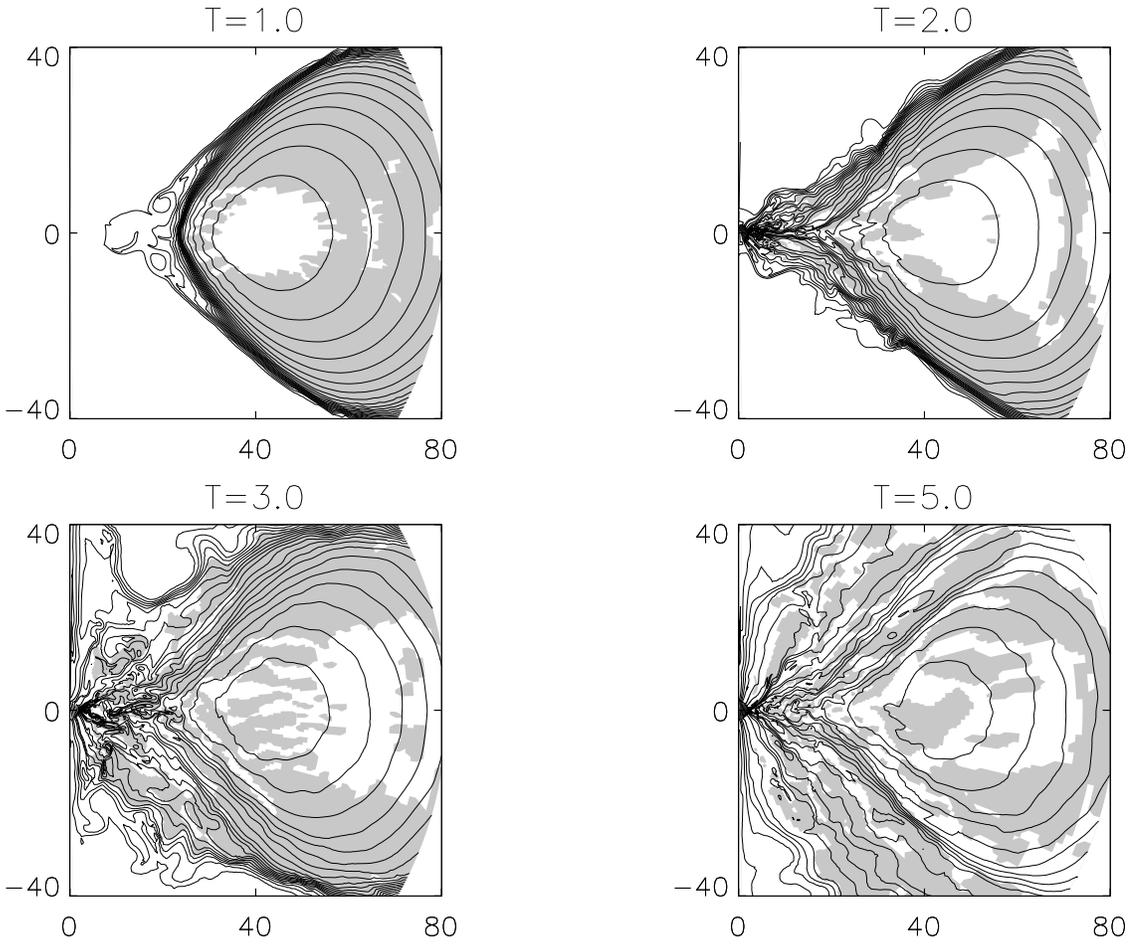}}
\end{picture}
\caption{Evolution of the density in a fiducial model (Run C).  Time is
measured in units of the orbital time at $r=40R_G$.  The axes are labeled
in units of $R_G$.  Twenty logarithmic contours are
used between the density maximum (0.97, 0.96, 0.94, and 0.85 for $t=1,2,3$
and 5 orbits respectively) and $10^{-4}$.
The shaded regions have $v_{r}>0$.}
\end{figure*}

The effect of numerical resolution on the nonlinear stage of the MRI
has been examined in a variety of local studies (e.g. HGB).
If the fastest growing mode is unresolved, saturation of the
instability occurs later and at a lower amplitude for the magnetic
energy.  Detailed analysis of the initial conditions in Run C show that
over most of the central regions of the torus $\lambda_c /
\bigtriangleup x \approx 6$ , where $\lambda_c$ is the wavelength
of the fastest growing
mode of the MRI given by $\lambda_c = 2\pi V_{A} / \sqrt{3} \Omega$ ($V_A$ and
$\Omega$ are the Alfv\'{e}n speed and orbital frequency respectively),
and $\bigtriangleup x$ is the grid size.
This indicates Run C is resolved, but just barely so.  Indeed, if Run C
is repeated at a lower resolution, collapse of the torus is delayed,
and fluctuations of, e.g. the density, are reduced.  Due to the increase in the
grid size with radius, and variations in the Alfv\'{e}n speed in the
torus, there are regions where the MRI is unresolved in Run C.  In order to
capture the evolution and interaction of many scales within the torus,
it will be important to increase the numerical resolution in future
studies beyond what is possible here.

Figure~2 is a plot of the time-history of the angle-integrated mass
accretion rate $\dot{M}_{\rm acc}$ through the inner boundary
in runs with different initial field strengths, where
\begin{equation}
  \dot{M}_{\rm acc}(r) =2\pi r^{2}\int_{0}^{\pi} \rho v_{r} \sin\theta d\theta.
\end{equation}
The data has been binned over 0.03 orbit intervals to construct the
plot, the instantaneous mass accretion rate shows even more rapid
fluctuations than are evident in Figure~2.
It is clear that at early times, the mass accretion rate depends on the
field strength: not surprisingly stronger fields have higher
$\dot{M}_{\rm acc}$.  Only Run B ($\beta_{\circ}=200$)
appears to reach a quasi-steady state;
in Run A ($\beta_{\circ}=100$) the accretion rate is highly variable,
while in Run C (the fiducial model, $\beta_{\circ}=400$) it shows
a slow but steady decline (although it seems to flatten out beyond orbit 4).
This decline is likely related to the
anti-dynamo theorem for axisymmetric flows; we discuss the differences
between these models in more detail in \S3.2.  Note that each
orbit at $r=R_{0}$ corresponds to nearly one hundred orbits
at the inner boundary, thus the decline in $\dot{M}_{\rm acc}$
in Run C occurs over many dynamical times for the inner regions
of the flow.  Shown as a dotted line
in Figure 2 is the mass accretion rate for a purely hydrodynamical
model computed using the Paczy\'{n}ski \& Wiita potential, an anomalous
shear stress of the form $\nu = 10^{-2}\rho$ (see SPB), and a grid
identical to that used for Runs A and B (hereafter we refer to this model
as Run D).  It shows a remarkably similar
profile to Run A, indicating that the magnetic stresses responsible for
accretion must have approximately the same amplitude as this
hydrodynamical model.  That is, the mass accretion rate in these MHD
models is small compared to what might be expected in an ADAF,
just as in the hydrodynamical models of SPB.

\begin{figure*}
\begin{picture}(252,180)
\put(0,0){\includegraphics{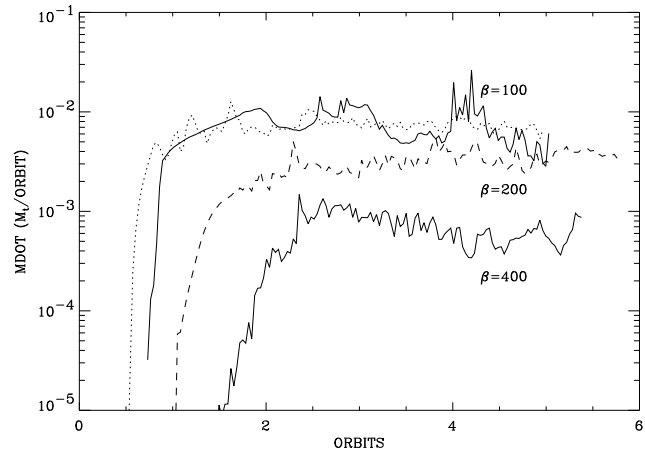}}
\end{picture}
\caption{Total mass accretion rate through the inner boundary for models with
different initial magnetic field strengths.  The solid and dashed lines labeled
$\beta=100$, $\beta=200$, and $\beta=400$ correspond to models A, B, and C
(the fiducial model)
respectively.  The dotted line shows the mass accretion rate for a purely
hydrodynamical model (Run D) with an anomalous shear stress $\nu=10^{-2}\rho$.}
\end{figure*}

Figure 3 plots the instantaneous structure of a number of quantities in
the inner regions $r<20R_{G}$ of Run C at $t=3.25$ orbits.  The flow is
clearly complex (and is also highly time-dependent).  Within $3R_{G}$
the profiles of each variable become radial as the gas free-falls
inward.  However, beyond this region each variable is dominated by
large amplitude fluctuations.  There is some anti-correlation between
the density and entropy fluctuation, although not as much as in the
purely hydrodynamical case (SPB).  This is an indication that the
velocity field which gives rise to the density fluctuations is not
driven by entropy gradients, i.e. that convection is not as
important in these MHD models (see also below).  That the entropy
gradients do not drive the flow in these simulations is also demonstrated
by setting the anomalous resistive heating rate to zero.  In this
case, we find turbulence and accretion continue despite the lack of
heating associated with dissipation of the magnetic field.  

The magnetic pressure associated with
the azimuthal component of the field is largest above and below the
midplane of the flow (and is in fact near zero in many regions close to
the midplane), which may be an indication of the effect of magnetic
buoyancy, or enhanced dissipation in regions of large current density.
The $r-\phi$ component of the Maxwell stress is highly
disordered in regions of the highest density (within $30^{\circ}$ of
the midplane).  This result is consistent with the saturated stage
of the axisymmetric modes of the MRI in fields with no net flux (Hawley
\& Balbus 1992).   

\begin{figure*}
\begin{picture}(553,287)
\put(0,0){\includegraphics{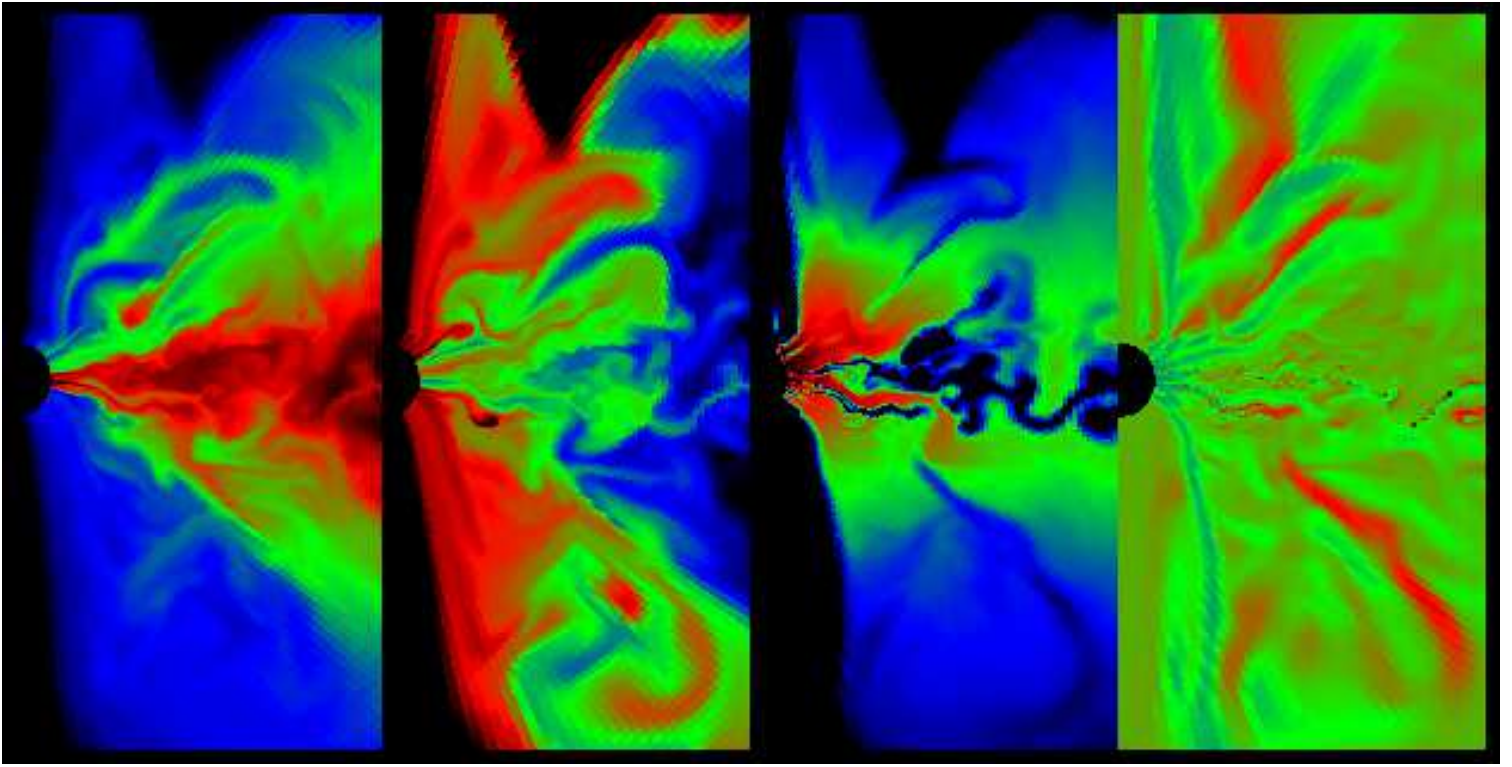}}
\end{picture}
\caption{(from left to right) Images of the logarithm of the density, specific
entropy, $B_{\phi}^{2}$, and $r-\phi$ component of the Maxwell stress normalized
by the magnetic pressure, i.e. $2B_{r}B_{\phi}/B^{2}$,
at $t=3.25$ orbits in the inner region
$r<20R_{G}$ of Run C.  The colour table runs from black to blue through red to
black.  The minimum and maximum of $\log \rho$ are -3.5 and -0.6,
of $S$ are -1 and 2, of $B_{\phi}^{2}$ are 0 and 3, and of
$2B_{r}B_{\phi}/B^{2}$ are -1.6 and 2.8.   }
\end{figure*}

In Figure 4 we plot two-dimensional contours of a large number of
quantities in Run C averaged between orbits 2.5 and 4.6.  Only 42 data
files have been used to construct this plot, thus many of the contours
are noisy.  Nonetheless, several important trends can be identified,
especially in comparison to the hydrodynamical case (Figure 4
of SPB).  Firstly, the contours of density and pressure are much more
similar in Run C than in the hydrodynamical models.  As a result, the
contours of the entropy $S$ are no longer nearly parallel to the contours
of the specific angular momentum $l=v_{\phi}r \sin \theta$,
instead the latter are more nearly
parallel to cylindrical radii than the former.  In a purely
hydrodynamical flow convection was found to be efficient enough to keep
the contours of $S$ and $l$ nearly parallel.
In the MHD flow the contours
of the angular frequency $\Omega$ are nearly parallel to cylinders,
whereas in the hydrodynamical case they were more spherical.  It would
appear that one action of the magnetic field is to try to reduce the
vertical gradients of $\Omega$.  In SPB it was argued that the ordering
of variables based on the shape of their contours was consistent with
marginal stability to the H{\o}iland criterion (Begelman \& Meier 1982;
Blandford \& Begelman 1999).  However, this ordering does not
apply to the MHD flow shown in Figure 4, confirming that even a weak
magnetic field changes the fundamental stability properties of a
rotating and stratified flow (Balbus 1995).

\begin{figure*}
\begin{picture}(504,360)
\put(0,0){\includegraphics{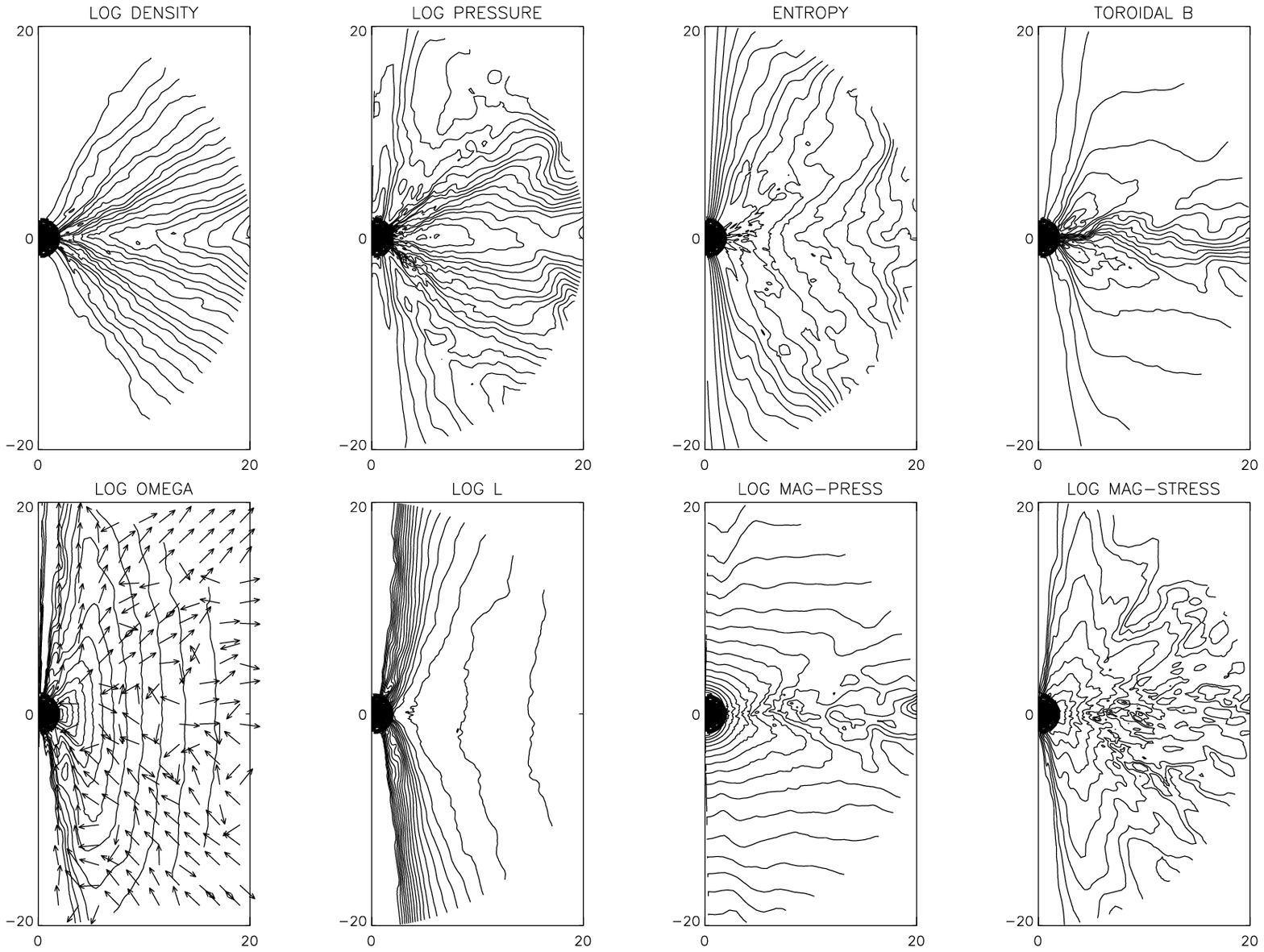}}
\end{picture}
\caption{Two-dimensional structure of a variety of time-averaged
quantities from the fiducial model (Run C).  The minimum and maximum of
$\log \rho$ are -3 and -0.71, of $\log P$ are -4.0 and -1.6,
of $S$ are -0.82 and 3.3, of $B_{\phi}$ are -0.19 and 0.21
of $\log \Omega$ are 0.48 and 2.1, of $\log l$ are -2 and -0.61,
of $\log B^{2}/8\pi$ are -3 and -0.61,
and of $\log B_{r}B_{\phi}/4\pi$ are -4 and -1.3.  Twenty equally spaced
contour levels are used, except $B_{r}B_{\phi}/4\pi$ for which ten are used.
The direction of the magnetic field at $t=3$~orbits is shown by unit vectors
plotted over the contours of $\Omega$}
\end{figure*}

Also plotted over the contours of $\Omega$ in Figure 4 are vectors showing
the direction of the poloidal magnetic field at $t=3$~orbits.  The
length of each vector has been scaled to unity, so that they indicate
the direction but not the magnitude of the field.  The vectors reveal
a highly tangled pattern in a wedge within $45^{\circ}$ of the equator,
consistent with the expected geometry of the field in MHD turbulence.
Towards the poles the field becomes much more ordered.  We use a snapshot
to show the field geometry rather than a time average because within the disk
the fluctuating field averages nearly to zero.
 
\subsection{Effect of variation of field strength}

In order to investigate the effect of the initial magnetic field
strength on the nature of the resulting accretion flow, we have
calculated two models identical to the fiducial model but
with $\beta_{\circ}=100$ and 200 respectively;
these are labeled Runs A and B in Table 1.  We have also calculated models
with the same initial magnetic field strength as Runs A and B, but which 
extend over a radial domain which is 2.5 times larger than the
fiducial model; these are labeled Runs E and F in Table 1.

The range of initial magnetic field strengths that can be studied
in the present set of experiments is limited by the numerical resolution
(which sets a lower limit) and the size of the torus (which sets an
upper limit).  Fields with $\beta_{\circ} < 100$ initially are too
strong for $\lambda_c$ to be smaller than the thickness of the torus,
we do not present results from such models here.
For models with $100 < \beta_{\circ} < 400$
global stresses from the radial component of the
field loops inside the torus produces magnetic stresses, outward
angular momentum transport, and inward accretion before the MRI
becomes nonlinear.  This is especially evident in Runs A and E, where
accretion begins after only about 0.7 orbits.  The accretion flow driven by
global stresses associated with a smooth field is very different from the
complex flow associated with accretion driven by the saturated MRI.
Figure 5 plots two-dimensional contours of a variety of quantities in Run F
averaged over orbits 2 through 3 using 20 data files.  Comparison
of Figures 4 and 5 reveal many differences.  The accretion flow in Run F
is confined to a thin, dense, wedge near the equatorial plane.
A strong and smooth toroidal field is located above and below the
midplane, strong vertical gradients in $B_{\phi}^{2}$ compress the
wedge.  Perhaps the only similarity between Figures 5 and 4 is that
once again contours of
$\Omega$ are parallel to cylinders.  Contours of the specific angular
momentum show that  the material slightly above the midplane
has a higher $l$ that material at the midplane (in the dense wedge there).
This evident as a small, outward bump in the contours of $l$ at the
midplane.

A snapshot of the direction of the magnetic field (but not its magnitude)
in Run F at $t=2$~orbits is shown by arrows plotted over the contours
of $\Omega$ in Figure 5.  Note that in contrast to the pattern
of the field lines in Run C (shown in Figure 4), at early times in
Run F the field is largely ordered throughout the flow.  Close to the
dense disk at the equatorial plane the field lines are highly inclined to
the vertical, but become more radial towards the poles.

The contours of $l$ in Figure 5 indicate that magnetic stresses
associated with an ordered poloidal magnetic field have resulted in a
deficit of angular momentum in material at the midplane, and an excess
of angular momentum in the material above it.  For this reason,
material above the midplane flows outward in a magnetocentrifugal
outflow (e.g., Blandford \& Payne 1981).  A small fraction of this
outflow near the poles is strongly unbound (with total energy $E>0$),
however most of the
denser outflow within $45^{\circ}$ of the midplane is still bound.  The
ratio of the mass outflow to accretion rate at the midplane is 0.1 at
$r=6R_{G}$, and increases only slightly to 0.2 by $r=50R_{G}$.  The
flow at this time bears a striking similarity to previous
two-dimensional global models of thin accretion disks (Shibata \&
Uchida 1986; Stone \& Norman 1994) threaded by a strong vertical
magnetic field.  Accretion and outflow in these thin disk simulations was
driven (for strong fields) by magnetic braking due to the effects of a
nonrotating corona in which the disk was embedded whereas in Run F, the
global stresses which drive accretion and outflow are associated with a
radial magnetic field that connects the inner and outer regions of the
torus.

Inevitably, growth and
saturation of the MRI in the denser regions of the torus, where the
radial field is initially smaller, perturbs and eventually overwhelms the
smooth accretion flow shown in Figure 5.
As a consequence, at late times (after orbit 3) Run F reverts to a
structure similar to Figure 4, that is a turbulent and rapidly
fluctuating flow driven by the MRI.

From Figure 2, it is clear that the mass accretion rate depends on the
initial magnetic field strength, with Run A having the highest $\dot{M_{acc}}$.
It is possible that even stronger vertical fields could drive
a higher $\dot{M_{acc}}$, a more
powerful outflow, and suppress the poloidal modes of the MRI.  On the other
hand, it is unlikely that such a configuration will be stable to 
the nonaxisymmetric MRI, thus 3D turbulent accretion dominated by the
MRI may be a more likely outcome.

\begin{figure*}
\begin{picture}(504,360)
\put(0,0){\includegraphics{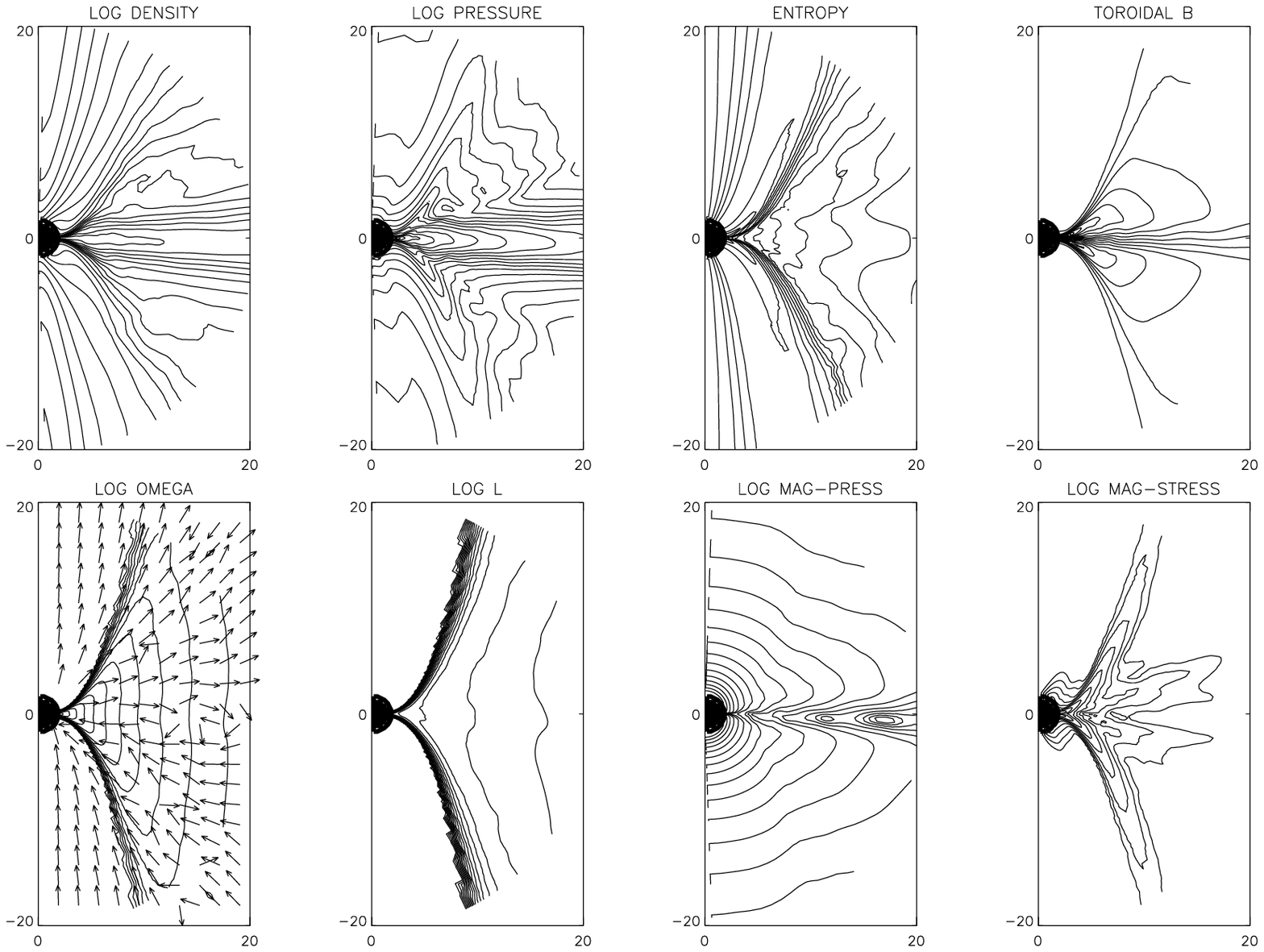}}
\end{picture}
\caption{Two-dimensional structure of a variety of time-averaged
quantities from Run F over orbits 2 through 3.  The minimum and maximum of
$\log \rho$ are -2.8 and -0.16, of $\log P$ are -2.9 and 0.79,
of $S$ are 0.63 and 2.3, of $B_{\phi}$ are -1.9 and 1.5,
of $\log \Omega$ are 0.48 and 2.6, of $\log l$ are -2 and -0.32,
of $\log B^{2}/8\pi$ are -1.9 and 1.6, and
of $\log B_{r}B_{\phi}/4\pi$ are -2 and 0.68.  Twenty equally spaced
contour levels are used for each variable except $B_{r}B_{\phi}/4\pi$,
for which 10 are used.
The direction of the magnetic field at $t=2$~orbits is shown by unit vectors
plotted over the contours of $\Omega$}
\end{figure*}

\subsection{Models over a larger radial domain}

Runs E through G listed in Table 1 are identical to models A, B, and D
respectively, except the density maximum of the torus is located at
$r=100R_G$ initially.  These models are considerably more expensive to
compute, because of the larger range in dynamical times between the
inner and outer regions of the grid.  Thus, we have not been able to
calculate the evolution of a $\beta_{\circ}=400$ model (which requires
128 grid points per decade in radius) over this large a radial domain.
Since these runs span a larger radial domain,
they are the more suitable for studying the radial
profiles of time-averaged variables in a MHD non-radiative accretion
flow.  On the other hand, because our models are axisymmetric, the flow
never becomes steady over dynamical times associated with the outer
regions of the disk.  Since there are roughly 350 orbits
of the disk near the inner boundary for every orbit at $r=R_{0}$ in these
models, however, the
flow may be quasi-steady in the inner regions.

Figure 6 plots the radial profiles of a variety of variables in Run F
angle-averaged over a small wedge near the midplane (between $\theta=84$
and 96 degrees), and time-averaged over 41 data files covering orbits
3 through 5.  This plot can be directly compared to Figure 5 in SPB.
The location of the last stable circular orbit is indicated by the
vertical dotted line in each panel.  Note that this same run was
shown at an earlier time in Figure 5, where it was noted that accretion
was dominated by global stresses rather than the MRI.  However,
by orbit 3 and for all times thereafter accretion of denser material 
driven by the nonlinear stage of the MRI dominates the flow.

\begin{figure*}
\begin{picture}(504,360)
\put(0,0){\includegraphics{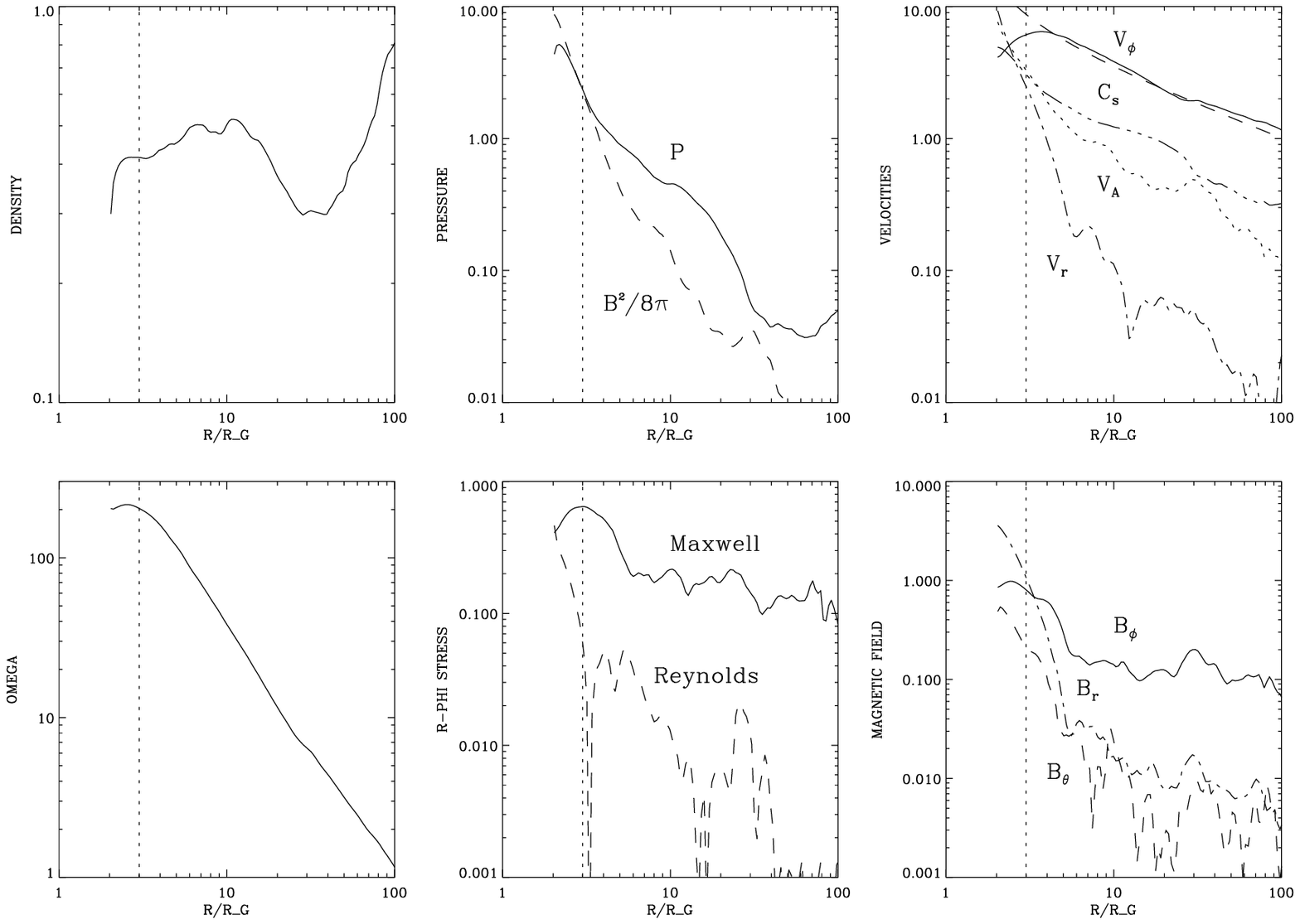}}
\end{picture}
\caption{Radial scaling of some time-averaged
quantities from Run F from orbits 3 through 5.  The solution is averaged over
angle between $\theta=84$ and 96 degrees to construct each plot.
The vertical dotted line in each plot shows the location of the last stable
circular orbit.
The dashed line in the plot of $v_{\phi}$ denotes Keplerian rotation
at the equator.  The Reynolds stress is calculated
using equation 9, only its amplitude is shown.}
\end{figure*}

In the hydrodynamical models of SPB, the profiles of each variable
became simple power-laws, whereas
in these MHD models the profiles are more complex.  For example, the
density is largely constant, although it increases by a factor of about
two near the outer edge.  The thermal and magnetic pressures are
identical at $r=3R_{G}$, and both decrease rapidly beyond roughly as
$r^{-3/2}$.  Note that the ratio of the gas to magnetic pressure ($\beta$)
is always greater than one
beyond $r=6R_{G}$.  The rotational velocity is nearly Keplerian
beyond $6R_{G}$, interior to this point it decreases rapidly.
Interestingly, the radial velocity increases sharply for $r<6R_{G}$ as
well, so that at $r=3R_{G}$ it is nearly equal to both the sound
and Alfv\'{e}n speed.
Rapid infall is therefore beginning well beyond the radius of the last
stable circular orbit.  Beyond $6R_{G}$ the decline in the radial
velocity is consistent with $r^{-1}$.

The magnetic stress, normalized to the magnetic pressure,
is nearly constant beyond $6R_{G}$ with $\alpha_{mag} = 2B_{r}B_{\phi}/B^{2}
\approx 0.1$.  (Note that $\alpha_{mag}$ is not the same as the Shakura-Sunyaev
$\alpha$ parameter.)
We measure the Reynolds stress using the difference
between the angular momentum flux and mass flux times the mean angular
momentum (Hawley 2000), i.e.
\begin{equation}
 \langle\rho v_{r}\delta v_{\phi}\rangle = \langle\rho v_{r}v_{\phi}\rangle
 - \langle \rho v_{r} \rangle \langle v_{\phi} \rangle
\end{equation}
This definition removes the uncertainty in the mean radial velocity
needed to compute the Reynolds stress from the fluctuating velocity
components alone.  Plotting the Reynolds stress in this fashion shows
that within $3R_{G}$ it grows rapidly in magnitude, however it is
negative, indicating angular momentum is being advected inwards by the
strong infall (note that only the magnitude of the Reynolds stress is
shown in Figure 6).  Beyond $3R_{G}$ the Reynolds stress is positive but
very small.

Finally, plots of each component of the magnetic field show that 
the $\phi$-component dominates over most of the radial domain.
Within $6R_{G}$, however, the radial component increases rapidly,
so that at $r=3R_{G}$, $B_{r} \approx B_{\phi}$.  This indicates the field
is being `combed' inwards by the rapid radial infall in this region.

Figure 6 reveals that the flow is strongly affected by the presence
of the last stable circular orbit well beyond $r=3R_{G}$.  In addition,
the large variations seen at large radii indicate the flow has
not reached a steady-state there (nor do we expect it to in axisymmetry).
This combination severely restricts
the radial domain over which any self-similar steady-state profiles
can be measured.  For the moment the profiles are consistent with 
$\alpha_{mag} = $constant, $v_{r} \propto r^{-1}$, and $\rho \propto$
constant.  However, these profiles must be measured in a fully 3D flow
that has reached a steady-state before any meaningful conclusions can
be drawn from them.

The profiles of the flow within $6R_{G}$ are in fact of great interest
in interpreting observations near the event horizon of
accreting black holes.
It is interesting that the large increase in the radial velocity occurs
{\em beyond} $3R_{G}$, presumably because of inward pressure gradients.
The inflow becomes supersonic just at $3R_{G}$, at which point
$B_{r} = B_{\phi}$.
These results are in agreement with recent models of MHD inflows near
the horizon of black holes developed by Krolik (1999) and 
Gammie (2000), even though such models are applicable to thin disks
as opposed to the thick disks studied here.
We also note the magnetic stress increases sharply within
$6R_{G}$, and is not zero at the last stable orbit.
The increase in the stress results from the
increase in $B_{r}$.  The rapid radial infall generates an
ordered radial component to the field, so that the $\langle B_{r}B_{\phi}
\rangle$ correlations are strongly increased.  In turn, the increased
stress affects the infall.
Note also from the plot of $\Omega$
in Figure 6 that $d \Omega / d \ln r$ is not zero at $3R_{G}$.
This means the flow is not strain-free at $3R_{G}$, so that if the
shear stress depends on the rate-of-strain, it will not be stress-free
there either.

If the nature of the flow near the inner region is set by the finite
pressure gradients associated with a thick disk, as opposed to magnetic
effects, similar profiles (e.g. rapid infall inside $6R_{G}$) should be
evident in a purely hydrodynamical simulation in a pseudo-Newtonian
potential.  Indeed, radial profiles of $v_{r}$ and $C_{s}$ in Run G
show a rapid increase in the infall velocity around $6R_{G}$, with the
flow becoming supersonic at $3R_{G}$, despite the fact that the shear stresses in the
hydrodynamical and MHD models depend on different quantities (e.g. the
former depends on the rate-of-strain).  Understanding the nature of
the infall region and how it changes with, e.g., the disk thickness is an
important problem for future studies.

By plotting the angle-integrated inward and outward mass flux rates,
$\dot{M}_{\rm in}$ and $\dot{M}_{\rm out}$ respectively,
defined through
\begin{equation}
 \dot{M}_{\rm in}(r) = 2\pi r^{2} \int_{0}^{\pi} \rho \min(v_{r},0)
   \sin \theta d\theta
\end{equation}
\begin{equation}
 \dot{M}_{\rm out}(r) = 2\pi r^{2} \int_{0}^{\pi} \rho \max(v_{r},0)
    \sin \theta d\theta
\end{equation}
SPB showed that in purely hydrodynamical flows driven by an anomalous
shear stress, $\dot{M}_{\rm in}$ increases rapidly with radius.  However,
the inward mass flux is
nearly exactly balanced by a large outward mass flux in the flow
associated with convective eddies, so that the net mass accretion rate
$\dot{M}_{\rm acc}$ defined in equation 8 is constant with radius.
The magnitude of the accretion rate $\dot{M}_{\rm acc}$ is set by the
properties of the flow at the inner boundary, and therefore is small
compared to the mass inflow rate $\dot{M}_{\rm in}$  at large radii.

To compare how the mass fluxes vary with radius in the MHD flows, in Figure 7
we plot $\dot{M}_{\rm in}$, $\dot{M}_{\rm out}$, and $\dot{M}_{\rm acc}$
for Runs E and F, that
is models with $\beta_{\circ}=100$ and 200 respectively.
In both cases, there is virtually no mass outflow
within $10R_G$.  Beyond $10R_G$, however, both the inward and
outward mass fluxes increase rapidly, whereas their sum remains
constant.  By the outer regions of the flow, the mass inflow rate
is two to four times the mass accretion rate.
It is not possible to measure the radial dependence of the increase
in the inward mass flux for these MHD models because of the limited 
dynamical range in radius.  

\begin{figure*}
\begin{picture}(504,180)
\put(0,0){\includegraphics{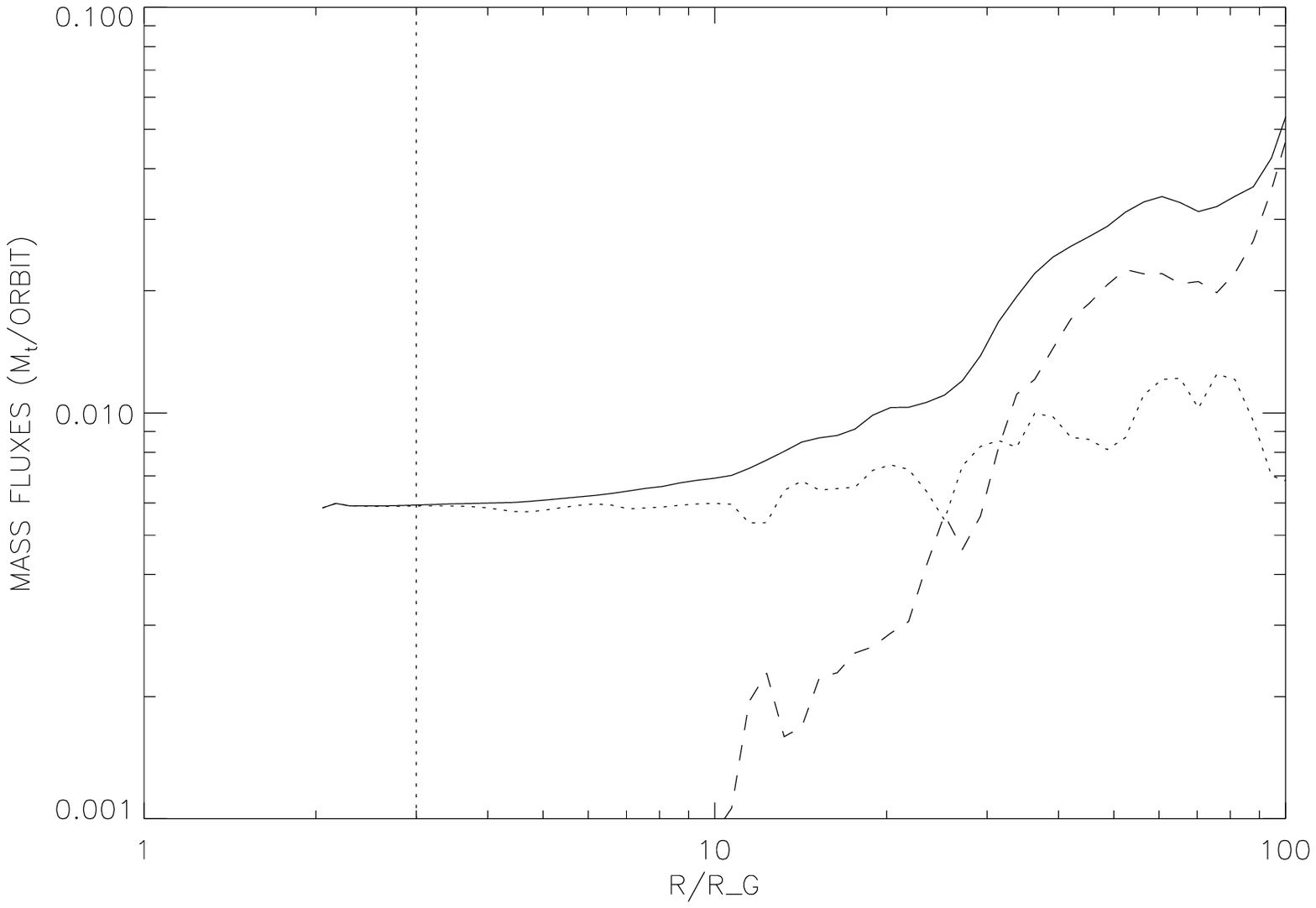}}
\put(252,0){\includegraphics{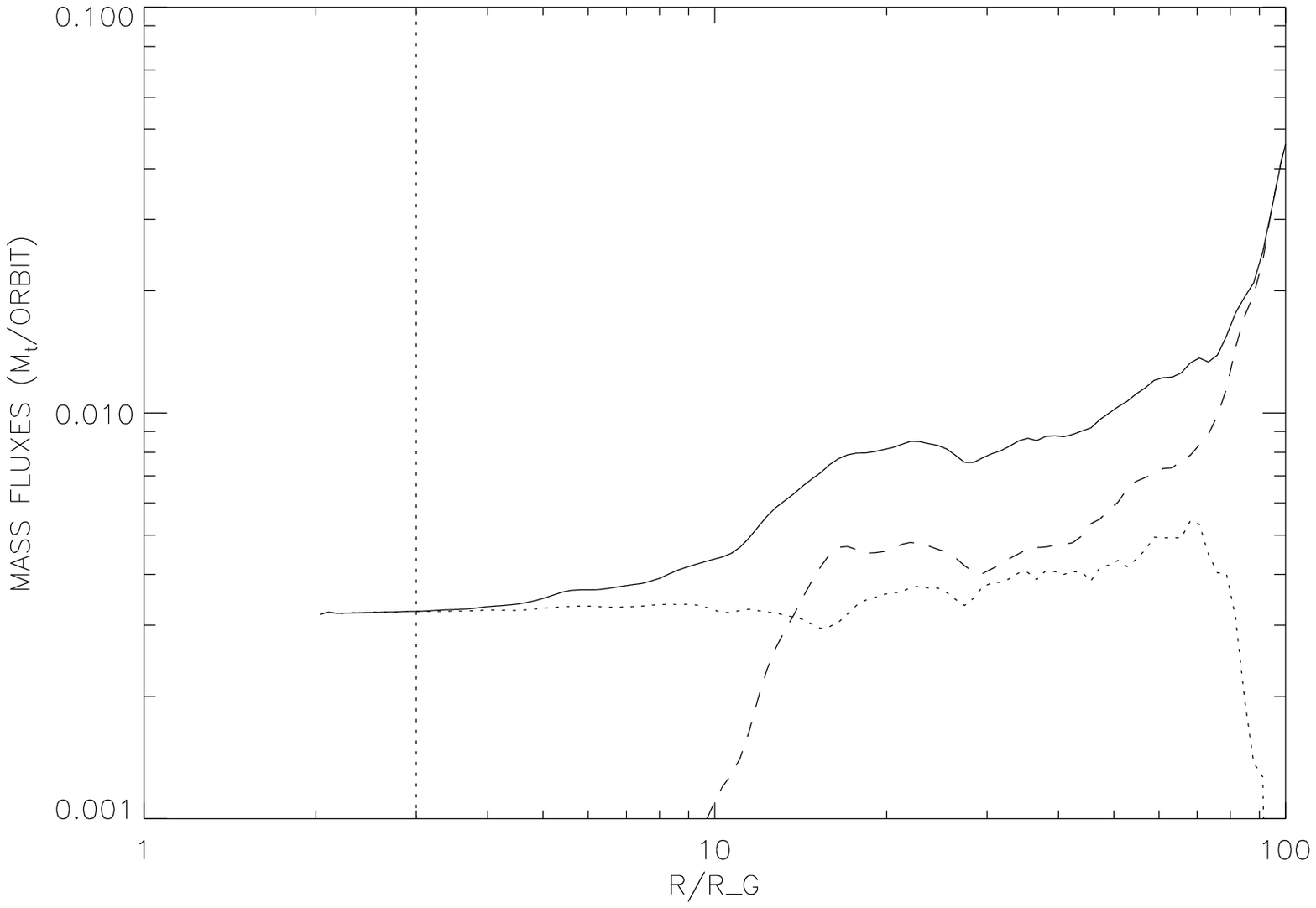}}
\end{picture}
\caption{Radial scaling of the amplitude of the  total mass inflow rate
$\dot{M}_{\rm in}$
(solid line), total mass outflow rate $\dot{M}_{\rm out}$ (dashed line), and
the net mass accretion rate  $\dot{M}_{\rm acc}$ (dotted line)
in Run E (left) and Run F (right).  These fluxes are defined by
equations 8, 10, and 11 in the text.  The vertical dotted line in each plot
shows the location of the last stable circular orbit.}
\end{figure*}

\section{Discussion}

The most important result of previous hydrodynamical simulations of
non-radiative accretion flows is that although the inward mass flux
associated with convective eddies is large, the net mass accretion rate
is small and set by the physical conditions at the inner boundary.  
Although this result was not expected from standard
self-similar solutions for ADAFs, analytic models
which reproduce the results of the simulations have been developed with the
ansatz that convection transports angular momentum inwards in axisymmetry
(Narayan, Abramowicz, \& Igumenshchev 2000; Quataert \& Gruzinov 2000).

The mass accretion rates observed in the two-dimensional MHD models
presented in this paper are consistent with the hydrodynamical
simulations, that is $\dot{M}_{\rm acc}$ is again small compared to the
expectations of a typical ADAF model.  However, in the case of MHD
models, turbulent eddies are not driven by convection but rather the
MRI.  In the MHD models there is no parameter that controls the
magnitude of the shear stress.  Instead, the magnitude of the Maxwell
stress scaled to the magnetic pressure $\alpha_{mag}$
results self-consistently from the dynamics.  Thus,
it is impossible to explore how this result varies with the amplitude of the
shear stress, as is typical for hydrodynamical models.  We find that
the small accretion rates reported here are produced even when
$\alpha_{mag} \approx 0.1 - 0.2$ (note that the Shakura-Sunyaev $\alpha$
parameter for these flows may be much different than $\alpha_{mag}$).
The amplitude of $\alpha_{mag}$ and
the mass accretion rates reported here are consistent with those given
by Hawley (2000) from fully three-dimensional simulations.  There is
growing observational support for small mass accretion rates around
black holes at the centers of nearby early-type galaxies, or the
galactic centre, (Di Matteo, Carilli, \& Fabian 2000; Agol 2000;
Quataert \& Gruzinov 2000), which provides encouragement for further
exploration of the dynamics reported here.

The large Alfv\'{e}n speed that can arise in regions of low density and
strong magnetic field makes these MHD models considerably more
expensive to compute that pure hydrodynamical simulations.  Thus, we
are not able to explore the dynamics over as large a radial domain as
used in SPB, even in axisymmetry, nor can we evolve them for as long.
Moreover, because of the lack of dynamo action to amplify the poloidal
field in axisymmetry, accretion in our models dies away after a few
orbits of the outer torus.  This makes our determination of the radial
scaling of the time- and angle-averaged flow ambiguous, but our results
are consistent with $v_{r} \propto r^{-1}$, and $\alpha_{mag} =$
constant.

In axisymmetry, we find accretion does not always occur because of
turbulent transport of angular momentum by the MRI.  For sufficiently
strong fields, global magnetic stresses associated with the radial
component of the field can drive outward angular momentum transport and
inward mass accretion.  In this case, we find unbound winds are
produced, as suggested by Blandford \& Begelman (1999),
but never observed in the hydrodynamical
simulations of SPB.  The mass flux carried in the outflow is small
compared to the mass accretion rate (typically the ratio of the two
is 0.1).  Moreover, the fraction of
outflowing material that is unbound is small, and it carries a small
fraction of the total outward angular momentum flux.  However, perhaps
stronger fields, or disks threaded by a net vertical flux would result
in more powerful MHD winds.  On the other hand, in 3D the
nonaxisymmetric MRI may destroy the smooth poloidal field structure and
therefore inhibit the production of winds.

By using the pseudo-potential of Paczy\'{n}ski \& Wiita, our
simulations reveal some interesting feature of the flow near the event
horizon of a black hole.  We find that infall begins at $r=6R_{G}$,
well beyond the radius of the last stable circular orbit.  We interpret
this due to the high pressure in a thick accretion flow.  The radial
infall becomes superAlfv\'{e}nic at $r=3R_{G}$.  The geometry of the
magnetic field is modified by the inflow, so that $B_{R} \sim
B_{\phi}$.  As a consequence the $r-\phi$ component of the magnetic
stress is increased significantly in the inner regions.  This stress is
non-zero at the radius of the last stable circular orbit.  All of these
effects are consistent with the expectations of recent analytic theory
(Krolik 1999; Gammie 2000)

Another evolutionary effect that is important in the inner regions of
the flow is the increase of net magnetic flux through the inner
boundary due to inward advection.  Although we study tori with zero net
magnetic flux, the inner regions (which carry one sign of vertical
field) are always accreted first.  The vertical field associated with
these regions is amplified by geometric compression, so that the
poloidal field can exceed the toroidal field near the inner boundary.
As the outer regions of the torus (which carry the opposite sign of
field) begin to accrete they can reduce the net flux through the inner
boundary.  Of course, this effect is strongly enhanced by axisymmetry.
It is not clear how large are the spatial correlations in a dynamo
amplified field in 3D.  However, because of the large range in spatial
and temporal scales, it is possible that a small patch in the outer
regions of the disk containing one sign of vertical flux may produce
similar behavior if it is accreted before the region containing the 
other sign of field.

We have focused on the time-averaged properties of the flow in our
simulations in order to compare to the results of steady-state theory.
However, there are a variety of interesting effects evident in the
instantaneous properties of the flow.  For example, Figure 8 is a
snapshot of the radial profile of the density and $r-\phi$ component of
the Maxwell stress at $t=2.35$~orbits in Run C, taken along the equator
($\theta=90^{\circ}$).  Extremely large fluctuations in the stress
(more than an order of magnitude) are evident, each of which is
associated with a large peak in the density.  These peaks represent
coherent blobs of matter which move inwards radially and accrete
periodically.  The profile of the stress indicates that the blobs are
formed by radial fluctuations in the stress.  In loose terms, this
behavior is analogous to the secular instability discussed by Lightman
\& Eardley (1974), although the assumption of axisymmetry may enhance
the effect in our simulations.  Such effects may imply the mass
accretion rate in an MHD flow will always be highly time-variable.

\begin{figure*}
\begin{picture}(252,180)
\put(0,0){\includegraphics{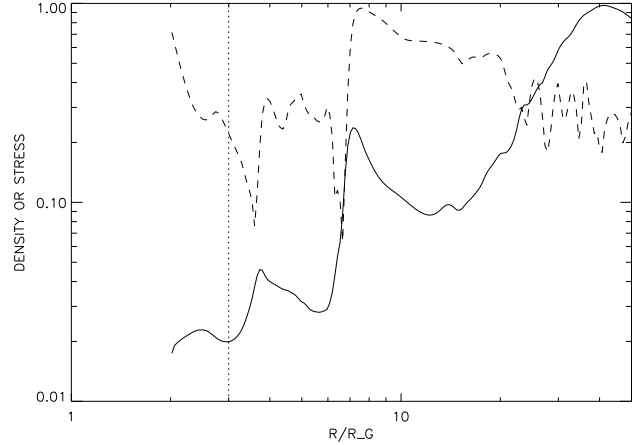}}
\end{picture}
\caption{Instantaneous radial profile of the density (solid line) and $r-\phi$
component of the Maxwell stress (dashed line) at $t=2.35$~orbits in Run C.
The vertical dotted line shows the location of the last stable
circular orbit.}
\end{figure*}

Finally, it is important to consider what are the limitations of this
current study.  Most importantly, we have studied only two-dimensional,
axisymmetric models.  This has several important consequences.
Firstly, we eliminate any nonaxisymmetric modes of the MRI, and
therefore limit the amplitude of the saturated state (and perhaps shear
stress).  Secondly, sustained dynamo action is not possible in
axisymmetry; on long timescales the MHD turbulence and accretion
eventually dies away in two-dimensional calculations.  This can be
dramatically demonstrated by a simulation which begins with the same
field geometry as studied here (field loops parallel to the density
contours), but with the direction of the field reversing on short
radial scales.  In this case, the field reconnects, and the turbulence
and accretion die away as soon as the MRI goes nonlinear.  The models
studied here are similar to flows with a net vertical flux in the sense
that only the inner regions of the torus (with one sign of field)
accrete on the timescale we have studied.  (As discussed above, this
also has the effect of driving an accumulation of net flux through the
inner boundary.) The importance of some of these effects can be gauged
by direct comparison of our results to the fully 3D simulations of
Hawley (2000).

In addition to being restricted to axisymmetry, our study is also
limited by the numerical resolution that is required for practical
calculations.  Since accretion is produced by a physical instability in
the flow that drives MHD turbulence, it is important to capture that
turbulence over a large range of scales.  It is clear that fully
three-dimensional simulations of MHD non-radiative accretion flows at
high resolution over a large radial domain are warranted, but will be
challenging.

\section{Conclusion}

We have presented the results of a series of time-dependent numerical
MHD simulations of the evolution of non-radiative accretion flows
around a black hole.  By confining this study to axisymmetry, we can
extend the calculations over a larger radial domain than is possible in
three-dimensions.  However, since sustained dynamo action is not
possible in axisymmetry, accretion in our simulations is transient, and
dies away after a few orbits of the outer disk.  Still, the large
dynamic range in radius in our simulations means that this represents
thousands of orbits of the inner regions of the disk; we have studied
the quasi steady-state flow that is produced in the inner disk during
this time.

There are a number of interesting similarities and differences in the
MHD flows studied here compared to previous hydrodynamical models (SPB,
IA99).

We find the net mass accretion rate is smaller than the mass inflow
rate at larger radii.  Over a large fraction of the accretion flow mass
inflow is balanced by mass outflow associated with turbulent eddies.
These eddies are driven by the MRI, rather than convection as in the 
hydrodynamical models.  This can be dramatically demonstrated in our
simulations by setting the resistive heating rate to zero; we find
turbulence and accretion driven by the MRI continues in this case.

The two-dimensional time-averaged structure of the flow is very different
than the hydrodynamical models.  Contours of the density and pressure
are nearly parallel, contours of the entropy and specific angular
momentum are not aligned, and the rotational frequency is constant on
cylinders.  It may be that the two-dimensional structure emerging in
these simulations is consistent with marginal stability to the
generalized H{\o}iland criterion in rotating, entropy-stratified, and
weakly magnetized flows (Balbus 1995), although further theoretical
work is required to confirm this.

The accretion flow is highly time-dependent.  A snapshot of the flow
shows that the amplitude of the Maxwell stress driving accretion, and
the density of the flow can fluctuate by an order of magnitude on the
local orbital timescale.

We find that accretion in these axisymmetric simulations proceeds both by 
turbulent flow driven by the MRI, and an ordered infall/outflow pattern
driven by global stresses associated with an ordered poloidal field.
The latter occur when the field is strong enough that the wavelength
of the fastest growing mode of the MRI is longer than the scale height of
the torus.  However, in 3D the nonaxisymmetric MRI may make it more
difficult to achieve the ordered infall/outflow pattern.

These simulations suggest that the enormous range in timescales between the
inner and outer regions of an accretion disk introduces complex dynamical
behavior if those
regions are coupled through a magnetic field.
Future three-dimensional simulations that span as large a radial
domain as used here promise to reveal much about the nature of
non-radiative accretion flows onto compact objects.

{\bf Acknowledgments:}
JS gratefully acknowledges financial support from the Institute of
Astronomy, and from NSF grant AST-9528299 and NASA grant NAG54278.

\appendix
\section{Artificial resistivity}

In order to convert the magnetic energy dissipated in current sheets
into thermal heating, we add an explicit artificial resistivity to the
induction equation.
The coefficient of artificial resistivity is given by
\begin{equation}
\eta = \frac{Q (\bigtriangleup x)^{2}}{\sqrt \rho} | {\bf J} |
\end{equation}
where $\bigtriangleup x$ is the grid spacing, and $Q$ is a
dimensionless constant.  This form is similar to that adopted by
Magara, Shibata, \& Yokoyama (1997) in studies of reconnection in
the solar corona.  The numerical implementation of the artificial
resistivity follows the technique outlined in Fleming, Stone, \& Hawley
(2000) for Ohmic dissipation, that is the resistive dissipation $\eta
{\bf J}$ is treated as an effective electromotive force used in the
constrained transport scheme to update the magnetic field components
(Stone \& Norman 1992b).

The amplitude $Q$ is specified by the requirement that the magnetic
Reynolds number $Re_{M} = \bigtriangleup x V_{A} / \eta$, where $V_{A}
= B/ \sqrt{4 \pi \rho}$ is the Alfv\'{e}n velocity, be of order unity
in current sheets in order for them to be spread out over a few zones.
Substituting the form for $\eta$ given above yields $Q=1/\sqrt{4 \pi}$
for $Re_{M} = 1$.  We have run several simulations with different
values of $Q$, keeping track of the evolution of the each component of
the total energy in the flow.  For $Q=0$ in Run B we find the change in
the total energy in the flow is 0.6\% after 2.7 orbits of evolution.
Using an artificial resistivity with $Q=0.1$ this change is reduced to
0.08\%, with most of the difference resulting from an increase in the
total thermal energy.  Thus, with an artificial resistivity,
conservation of total energy is improved because energy losses due to
numerical reconnection are reduced, and instead magnetic energy is
converted directly to thermal energy in current sheets.  An explicit
artificial resistivity as introduced here could be important for any
flow in which heating due to magnetic reconnection is important that is
studied using a numerical algorithm that solves the internal (rather
than total) energy equation.

\end{document}